\documentclass{JAC2000}

\usepackage{graphicx}

\begin{document}
\title{JLC PROGRESS}

\author{N.~Toge, KEK, Tsukuba, Ibaraki 305-0801, Japan}

\maketitle

\begin{abstract} 
The JLC is a linear collider project pursued in Japan by researchers 
centered around KEK. The R\&D status for the JLC project is presented, 
with emphasis on recent results from ATF concerning studies of 
production of ultra-low emittance beams and from manufacturing 
research on X-band accelerator structures.
\end{abstract}

\section{Introduction}

Major elements of the current R\&D activities for the JLC 
project\cite{JLC} includes: development of polarized electron 
sources\cite{Nakanishi}, experimental studies of a damping 
ring\cite{ATFdesign}, development of X-band technologies as the main 
scheme for the main linacs, and C-band RF development as a backup 
technology for the main linacs\cite{C-band}.

\begin{figure*}[!htb]
\centering
\includegraphics*[width=165mm]{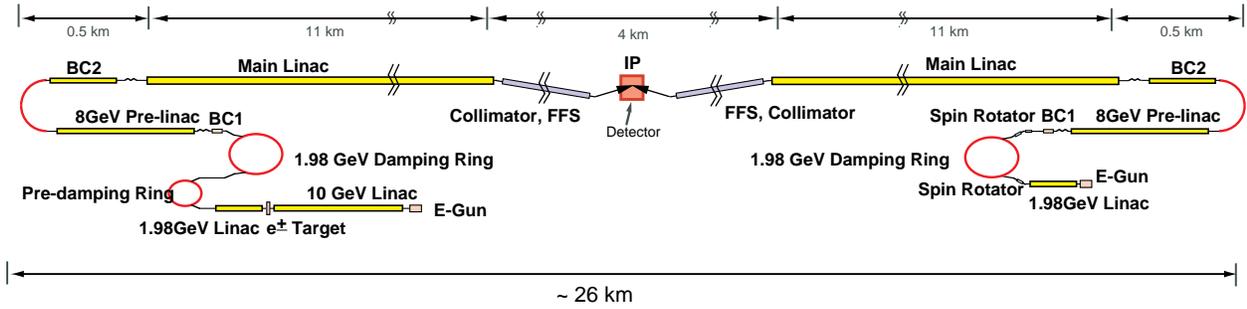}
\caption{Schematic layout of JLC in its $E_{cm} = 1$~TeV configuration.
\label{JLC_schematics}}
\end{figure*}

Figure\,\ref{JLC_schematics} shows a schematic diagram of JLC. The 
target center-of-mass energy is 250$\sim$500 GeV in phase-I, and 
$\sim$1 TeV or higher in phase-II. Since 1998, through an R\&D 
collaboration (International Study Group -- ISG) which was formalized 
between KEK and SLAC\cite{ISG}, development of hardware elements for 
the X-band main linacs has been pursued based on the basic parameters 
common to both JLC and NLC\cite{NLC}.  Tables\,\ref{X_table_1} and 
\ref{X_table_2} give the most up-to-date basic machine parameters that 
have been chosen as a result of optimization process in ISG discussions.

\begin{table}[!hhh]
\begin{center}
    \caption{Partial list of representative JLC parameters as of April, 
2000
\cite{ISG}, if the main linacs are built based on the X-band technology.
\label{X_table_1}}
\begin{tabular}{lll}
\hline\hline
Item & Value & Unit \\
\hline
\#Electrons / bunch & 9.5\(\times\)10$^9$ & \\
\#Bunches / train & 95 & \\
Bunch separation & 2.8 & ns \\
Train length & 263.2 & ns \\
RF frequency & 11.424 & GHz \\
RF wavelength & 26.242 & mm \\
Klystron peak power & 75 & MW \\
Length / cavity unit & 1.8 & m \\
\(a/\lambda\) & average 0.18  & \\
Filling time & 103 & ns \\
Shunt impedance & 90 & M\(\Omega\)/m \\
\(E_{acc}\)(no-load) & 72 & MV/m \\
\(E_{acc}\)(loaded) & 56.7 & MV/m \\
Normalized emittance & 3.0 $\times$ 0.03 (Linac) & 10$^{-6}$m.rad \\
 & 4.5 $\times$ 0.1  (IP)  & 10$^{-6}$m.rad \\
Bunch length & 120 & \(\mu\)m \\
\hline
\end{tabular}
\end{center}
\end{table}

\begin{table}[!hhh]
\begin{center}
    \caption{Representative JLC parameters (continued), if the main linacs are 
built based on the X-band technology. Parameters that would vary for 
$E_{\rm CM} = $500~GeV and 1.0~TeV are shown. 
\label{X_table_2}}

\begin{tabular}{llll}
\hline\hline
\(E_{\rm CM}\) & 500~GeV & 1~TeV & \\
\hline
\#cav/linac & 2484 & 4968 & \\
\#klystrons/linac & 1584 & 3312 & \\
Length/linac & 4.3 & 8.9 & km \\
P(wall-plug) & 94 & 191 & MW \\
Rep. rate & 120 & 120 & Hz \\
\(\beta_x^*\times\beta_y^*\) & 12$\times$0.12 & 12$\times$0.15 & mm\(\times\)mm
\\
\(\sigma_x^*\times\sigma_y^*\) & 330$\times$4.9 & 235$\times$3.9 & nm \(\times\)
nm \\
\(<-\Delta E/E>\) & 4.0 & 10.3 & \% \\
due to BSM & & & \\
Lum. pinch  & 1.1 & 1.43 & \\
enhancement & & & \\
Luminosity & 7$\times$10$^{33}$ & 13$\times$10$^{33}$ & cm\(^{-2}\)s\(^{-1}\) \\
\hline
\end{tabular}
\end{center}
\end{table}

This paper 
presents the R\&D status of the JLC project, with strong focus on (i) the 
most recent results from ATF concerning studies of production of ultra-low 
emittance beams and (ii) manufacturing research on X-band accelerator 
structures.

\section{ATF}

\begin{figure*}[!htb]
\centering
\includegraphics*[width=135mm]{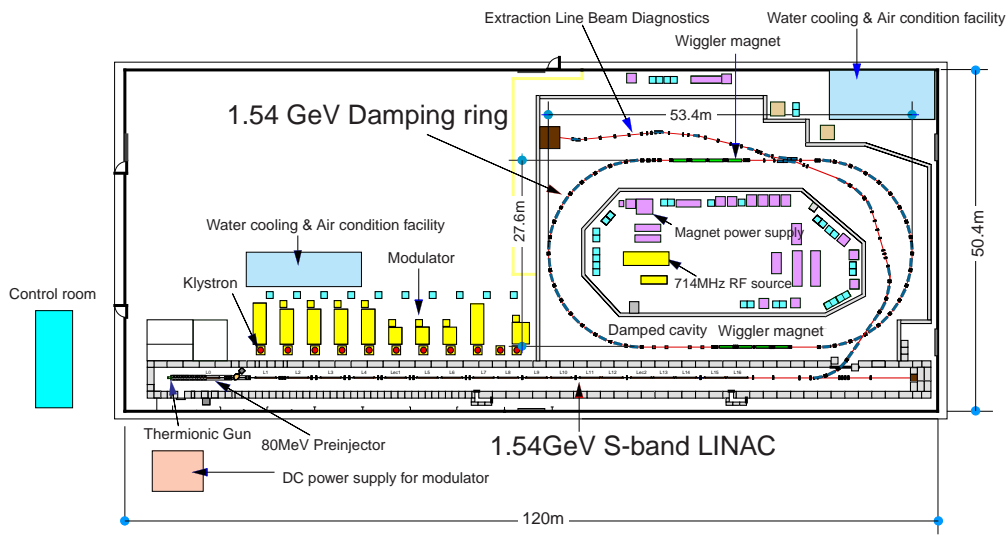}
\caption{Layout of ATF -- Accelerator Test Facility -- at KEK.}
\label{ATFscheme}
\end{figure*}

\begin{table*}[!htb]
    \centering
            \caption{Achieved and design parameters at ATF.}
        \label{ATFparam}
    \begin{tabular}{lccl}
        \hline\hline
        Item & Achieved & Design & Unit \\
        \hline
        Linac Status & & & \\
        Max. beam energy & 1.42 & 1.54 & GeV \\
        Max. gradient with beam & 28.7 & 30 & MeV/m \\
        Single bunch population & 1.7$\times$10$^{10}$ & 
        2$\times$10$^{10}$ & \\
        Multi-bunch population & 7.6$\times$10$^{10}$ & 40 
        $\times$10$^{10}$ & \\
        Bunch spacing & 2.8 & 2.8 & ns \\
        Repetition rate & 12.5 & 25 & Hz\\
        Energy spread (full width) & $<$ 2.0 \% (90\% beam) & $<$ 1.0 \% (90\% 
         beam) & \\
        \hline
        Damping Ring Status & & & \\
        Max. beam energy & 1.28 & 1.54 & GeV \\
        Circumference & 138.6 $\pm$ 0.003 & 138.6 & m \\
        Momentum compaction & 0.00214 & 0.00214 & \\
        Single bunch population & 1.2$\times$10$^{10}$ & 
        2$\times$10$^{10}$ & \\
        COD (peak-to-peak) & $x\sim$2, $y\sim$1 & 1 & mm \\
        Bunch length & $\sim$6& 5 & mm \\
        Energy spread & 0.06 \%& 0.08 \% \\
        Horizontal emittance & (1.4$\pm$0.3)$\times$10$^{-9}$ & 
        1.4$\times$10$^{-9}$ & m \\
        Vertical emittance & (1.5$\pm$0.25)$\times$10$^{-11}$ & 
        1.0$\times$10$^{-11}$ & mm \\
        \hline
        \end{tabular}

\end{table*}

The Accelerator Test Facility (ATF)\cite{ATFdesign} at KEK (see 
Figure~\ref{ATFscheme}) is a test bed for an upstream portion of JLC, 
which has to produce a train of ultra-low emittance bunches of 
electrons (and positrons).  It includes a multi-bunch-capable electron 
source, a 1.54\,GeV S-band linac, and a 1.54\,GeV damping ring (DR) 
prototype.

A variety of studies were successfully conducted in 1994 through 1996 
on acceleration of multi-bunch beam (up to 12 bunches, 2.8\,ns bunch separation).  The multi-bunch beam 
loading compensation scheme based on the RF frequency modulation of a 
small set of accelerating structures was successfully demonstrated.

Commissioning work of the ATF DR began in early 1997 with many 
participants from both inside and outside KEK. The accelerator has been 
operated in a single-bunch mode with the typical stored intensity of 
\(\sim1\times10^{10}\) electron/bunch or less at a repetition rate up 
to 1.56 Hz.  Achieved and design parameters relevant to operation of 
ATF are summarized in Table~\ref{ATFparam}.  By Summer 1998, the 
horizontal beam emittance $\epsilon_{x}$ of 
$\sim$1.4$\times$10$^{-9}$m (i.e. $\gamma\epsilon_{x} \simeq$ 
3.5$\times$10$^{-6}$ m) was measured by using a group of wire scanners 
in the beam diagnostics section of the extraction line\cite{Xemit}.  
\clearpage

\begin{figure}[htb]
\centering
\includegraphics*[width=60mm]{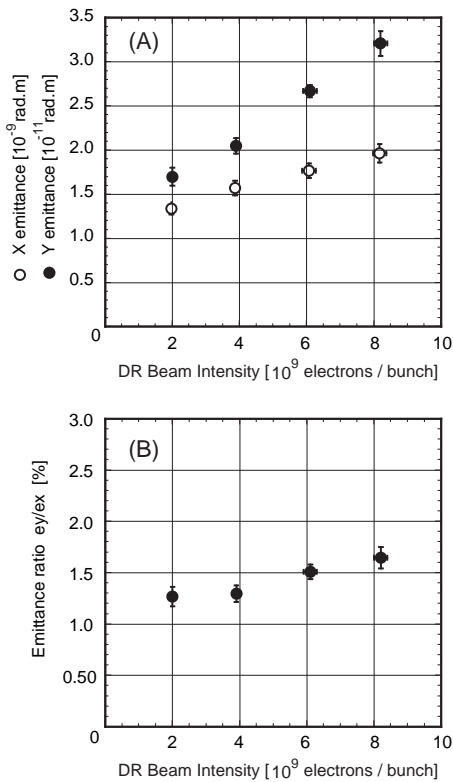}
\caption{Measured values of (A) horizontal and vertical beam 
emittance (unnormalized) and their ratio (B) as function of bunch 
intensity from ATF.}
\label{DRemit}
\end{figure}

The most recent set of measured horizontal and vertical emittance 
values\cite{DRstatus}, as of April, 2000, are shown in 
Figure~\ref{DRemit}.

Some of the development at ATF which are considered to play important 
roles in the progress of the achieved vertical emittance values are 
summarized as follows:
\begin{itemize}
    \item Improved resolution ($\sim$20 $\mu$m) of single-shot BPM
    readout electronics.
    \item Improved understanding of the first order optics in the DR, and 
    corrections introduced by using ``fudge'' factors for the field 
    strength of quadrupole magnets and quadrupole field components of 
    the combined-function bend magnets in the DR.
    \item Improved dispersion and orbit correction algorithm which are 
    tuned to minimize the vertical dispersion ($\sim$ 5 mm) in the DR without 
    overly upsetting the COD.
    \item Skew quadrupole magnet fields were introduced in the arc 
    sections of the DR by using trim windings of sextupole magnets. 
    They were used to minimize the cross-plane coupling by using the 
    $x$-$y$ coupling signals in the diagnostics of the first-order 
    optics, as well as by using the tune difference $\nu_{x}$-$\nu_{y}$ 
    near the coupling resonance. 
    \item Improved algorithm for the correction procedure for the 
    vertical dispersion in the extraction line, where the wire scanner
    beam diagnostics instruments are situated.
\end{itemize}

\begin{figure}[htb]
\centering
\includegraphics[width=60mm]{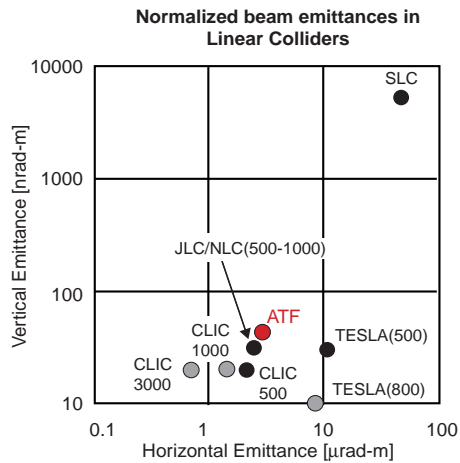}
\caption{Normalized horizontal and vertical emittance values that 
have been achieved 
at SLAC and ATF, compared to what are required for injectors of next-generation 
linear colliders.}
\label{emitsummary}
\end{figure}
This result from ATF may be put in perspective as shown in 
Figure~\ref{emitsummary}. It is seen that the beam emittance that is 
required for typical next-generation linear colliders, including 
JLC/NLC, is nearly achieved. However, a number of issues still remain 
to be investigated at ATF. For instance,
\begin{enumerate}
    \item The beam emittance values so far considered the most 
    reliable have been obtained by using wire scanners in the 
    extraction line. These and measurements from synchrotron 
    radiation (SR) monitor in the DR, which utilizes the interferometry, 
    are not totally consistent. This is most likely due to effects of 
    mechanical vibrations of the optical stands that are used for the 
    SR monitor system, but it requires more studies.
    \item There may be field errors in the magnetic components in the 
    beam extraction line or at the beam extraction point. They  might 
    introduce $x$-$y$ cross plane coupling and fictitious  signals 
    of the growth of the vertical beam emittance, which may not yet be
    accounted for in wire scanner measurements.
    \item Observed growth of vertical emittance or the emittance ratio 
    as shown in Figure~\ref{DRemit} is found to have too strong a 
    dependence on the bunch intensity, compared to existing model calculations 
    of intra-beam scattering effects. It has not yet been resolved 
    whether this is due to errors in measurements, inadequate set-up assumptions 
    or true beam dynamics effects.
    \end{enumerate}
In addition, the reported emittance numbers from ATF are so far based 
only on single-bunch beam operations.  After the Summer shutdown 
period, the beam operation of ATF is scheduled to resume in October, 
2000. Preparation is currently under way for addressing the single-bunch 
emittance issues as well as multi-bunch operation of the ATF DR.

\section{X-band Accelerating Structure}

Development of X-band accelerating structure at KEK has been conducted 
in close collaboration with a group at SLAC. The accelerating 
structure studied is based on the damped-detuned concept~\cite{ISG}.  The 
recent research focus at KEK has been on (i) fabrication of copper disks 
for the RDDS (Rounded Damped-Detuned) structure with a diamond-turning 
technique with ultra-high precision lathes, and (ii) their assembly into 
structure bodies by means of the diffusion bonding technique\cite{ISG, 
RDDS1}. They have been pursued in conjunction with development of 
better control of transverse wakefield and improved RF-to-beam 
efficiency.

\begin{figure}[!htb]
\centering
\includegraphics*[width=65mm]{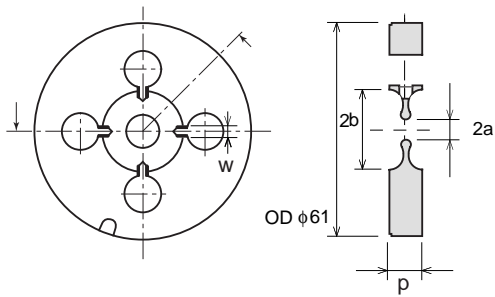}
\caption{Schematic drawing of an RDDS disk.}
\label{RDDSdisk}
\end{figure}

Figure~\ref{RDDSdisk} shows a schematic drawing of the typical copper 
disk for the RDDS structure, whose first prototype (RDDS1) was 
successfully built in 1999 and was tested for wakefield 
characteristics at the ASSET facility of SLAC in 2000.  The RDDS1 (1.8 
m long) consists of 206 similar disks, each 61 mm in diameter and 
8.737 mm in thickness.

\begin{figure}[!htb]
\centering
\includegraphics*[width=65mm]{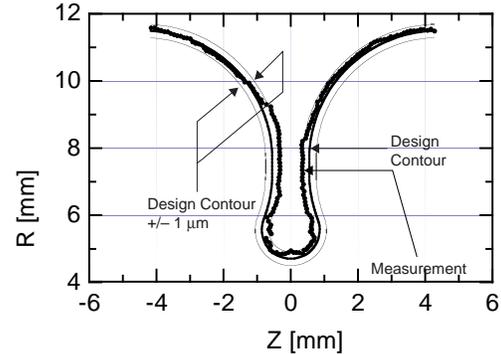}
\caption{Contour profile plot of a test RDDS disk near the aperture 
opening part.}
\label{RDDScontour}
\end{figure}

During disk fabrication, much attentions have been paid to the 
temperature control of the lathe, positioning of the diamond cutting 
tool, and its motion.  The radius of the cutting tool are 
pre-determined within 0.3 $\mu$m by machining an aluminum test 
hemisphere ($\phi$60 mm) and by measuring the surface features with a 
roundness tester.  Figure~\ref{RDDScontour} shows a contour profile 
plot of a test RDDS disk that was measured with a CMM (Coordinate 
Measurement Machine) with contouring capability.  The solid line shows 
the design shape, while the black dots show the measured shape, with the 
deviation from the nominal shape magnified by 200 times.  
The machined surface matched the design to within $\pm$1 $\mu$m.  
Other dimensional parameters such as the disk outer diameter, 
aperture radius $2a$, which are determined by the diamond 
turning, are found to have been done with similar precision.


\begin{figure}[!htb]
\centering
\includegraphics*[width=75mm]{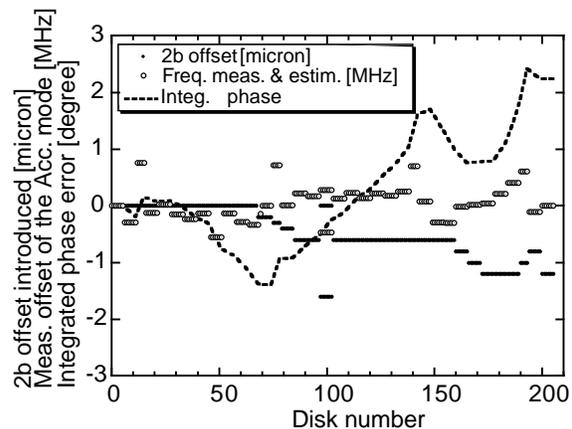}
\caption{Fundamental mode frequencies, measured in a disk-stack setup,
of RDDS1 disks.}
\label{RDDSrf}
\end{figure}

In addition to mechanical quality control, careful measurements of the 
fundamental and first-order transverse mode resonant frequencies of 
cells were performed for individual disks.  As an example, 
Figure~\ref{RDDSrf} shows the fundamental-mode frequencies, measured 
on a disk-stack setup (white circles).  The disks are mostly 
fabricated in the order of the disk number.  Figure~\ref{RDDSrf} also 
shows the integrated phase error (broken line), expected from the 
measured deviation of the fundamental-mode frequencies.  It is seen 
that by fine-adjusting the cell cavity size (denoted as $2b$ in 
Figure~\ref{RDDSrf}) based on the past trend of frequency errors, the 
total integrated phase error can be controlled with an extreme 
precision.  Also, the first-order transverse mode frequencies of the 
fabricated disks have been found to have a smooth distribution within 
0.4 to 0.6 MHz.  Overall, the micron-level precision to which the 
disks are machined have been very successfully demonstrated.

The disks are then stacked on a precision V-block, where a disk-to-disk 
alignment of 1 $\mu$m or better is possible during stacking.  
A growth of the so-called ``bookshelf'' stack errors is prevented 
by monitoring the inclination angles of the surface of stacked disks 
with a two-axis autocollimator and by applying corrections during 
stacking.

Formation of the complete accelerating structure is made through (i) 
the diffusion bonding procedure of the copper disks that is conducted 
in two steps (prebonding and final bonding), and (ii) the brazing of 
external components such as cooling water tubing, fixtures for support 
frames, waveguides and flanges.  It has been found repeatedly that the 
disk-to-disk alignment and the ``bookshelving'' error (of its 
absense thereof) of the disks are well maintained throughout the 
diffusion bonding process, which create vacuum-tight and mechanically 
strong enough disk-to-disk bonding junctions.

However, differential expansion of the ceramics endplate supports 
relative to the first and the last copper disks lead to a flaring of 
the structure ends during diffusion bonding.  Similar deformation 
occurred on RDDS1 when a stainless steel manifold was installed on a 
mid portion of the structure during brazing.  While the former error 
could be rectified later, the bonding techniques used in the assembly 
of X-band structures call for some improvements in the near future, in 
addition to studies of mass-production issues.  Results from wakefield 
measurements of RDDS1 prototype are reported in a contribution 
submitted to this conference~\cite{RDDS1}.  Also, issues pertaining to 
RF processing and high power operation of X-band accelerating 
structures at field gradient up to 70$\sim$80 MV/m are being investigated\cite{RFproc}.

\section{Other Activities on the X-band RF R\&D}

\begin{figure}[!htb]
\centering
\includegraphics*[width=75mm]{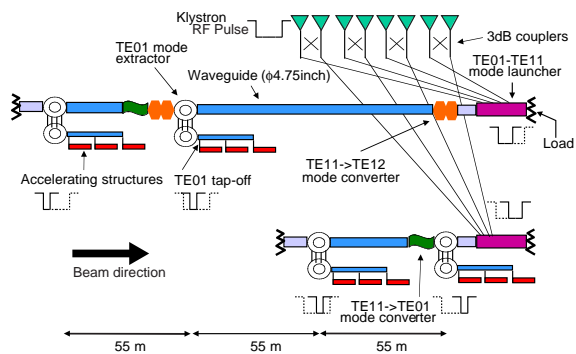}
\caption{Schematic diagram of a DLDS concept where the RF power from 8 
klystrons are divided and distributed to four clusters of accelerating 
structures situated along the linac.}
\label{DLDS}
\end{figure}

Development work is also under way\cite{Xstatus} for: klystron modulators with 
semiconductor switching devices, construction of X-band high-power 
klystrons with periodic-permanent magnet (PPM) focusing, testing of 
the DLDS (Delay Line Distribution System) components (see 
Figure~\ref{DLDS}), development of X-band 
high-power RF windows\cite{THA02}.  Some of the efforts are carried 
out in collaboration with a group from Protvino branch of BINP, 
Russia, as well as with SLAC in the framework of the KEK-SLAC ISG.

A low-power testing of transmission of X-band RF through a long 
($\sim$50 m) waveguide as a proof-of-principle experiment of the DLDS 
concept was successfully conducted at KEK by a KEK-SLAC-Protvino 
collaboration in 1999.  Another testing is planned for Fall, 2000.  A 
high-power testing of Protvino-KEK RF windows have been successfully 
carried out at SLAC in late 1999 through early 2000.  As of August, 
2000, intense testing is being conducted on KEK site for a PPM 
klystron that was designed at KEK and built in collaboration with 
Japanese industry.

In a few years, when the basic R\&D of these RF components, 
including the accelerating structures, becomes sufficiently mature, it 
is considered highly desirable to build a small part of the  
X-band linac, for instance a complete unit set of the RF system 
as shown in Figure~\ref{DLDS}.  While its successful operation without any beam 
acceleration should already mark a major milestone, possible 
acceleration of low-emittance beams which would be hopefully available 
by that time from ATF would play a decisive role in showing the 
feasibility of JLC/NLC.

\end{document}